\documentclass[a4paper,10pt,preprint,5p]{elsarticle}
\usepackage{amsmath}
\usepackage{amssymb}
\usepackage{amsfonts}
\usepackage{bm}

\biboptions{sort&compress}
\usepackage{gensymb}
\usepackage{lipsum}
\usepackage[]{graphicx}
\usepackage{color}
\usepackage[]{caption}
\usepackage[]{subcaption}

\usepackage[unicode, pdftex]{hyperref}
\usepackage{lipsum}

\begin{document}
\begin{frontmatter}
\title{Simple model for the gap in the surface states of the antiferromagnetic topological insulator MnBi$_2$Te$_4$}
	
\author[1,2,3]{R. S. Akzyanov\corref{cor1}}
\ead{akzyanov@phystech.edu}

\author[1,3]{A. L. Rakhmanov}

\cortext[cor1]{Corresponding author.}
    
\address[1]{Dukhov Research Institute of Automatics, Sushchevskaya 22 st., 127030, Moscow, Russia}
    
\address[2]{Moscow Institute of Physics and Technology, Dolgoprudny, Moscow Region, 141700 Russia}

\address[3]{Institute for Theoretical and Applied Electrodynamics, Russian Academy of Sciences, Moscow, 125412 Russia}

\begin{abstract}
We study the influence of the antiferromagnetic order on the surface states of topological insulators. We derive an effective Hamiltonian for these states, taking into account the spatial structure of the antiferromagnetic order. We obtain a typical (gapless) Dirac Hamiltonian for the surface states when the surface of the sample is not perturbed. Gapless spectrum is protected by the combination of time-reversal and half-translation symmetries. However, a shift in the chemical potential of the surface layer opens a gap in the spectrum away from the Fermi energy. Such a gap occurs only in systems with finite antiferromagnetic order. We observe that the system topology remains unchanged even for large values of the disorder. We calculate the spectrum using the tight-binding model with different boundary conditions. In this case we get a gap in the spectrum of the surface states. This discrepancy arises due to the violation of the combined time-reversal symmetry. We compare our results with experiments and density functional theory calculations.
\end{abstract}
\end{frontmatter}
\section{Introduction}

Magnetic topological insulators (MTIs) are narrow-gap semiconductors that exhibit a nontrivial band structure along with magnetic order~\cite{Tokura2019,Wang2021r,Bernevig2022}. A prominent feature of the topological insulators (TIs) is the presence of the surface states that are robust against disorder~\cite{Hasan2010, Fiete2012}. The exchange interaction in the MTIs breaks the time-reversal symmetry of the system and can open a band gap in the spectrum of the surface electron states~\cite{Liu2010,Chang2023}. This significantly distinguishes MTIs from non-magnetic TIs and makes it possible to observe the anomalous quantum Hall effect and chiral Majorana states~\cite{He2020,Li2023}. 

\begin{figure}[ht]
\includegraphics[width=1\linewidth]{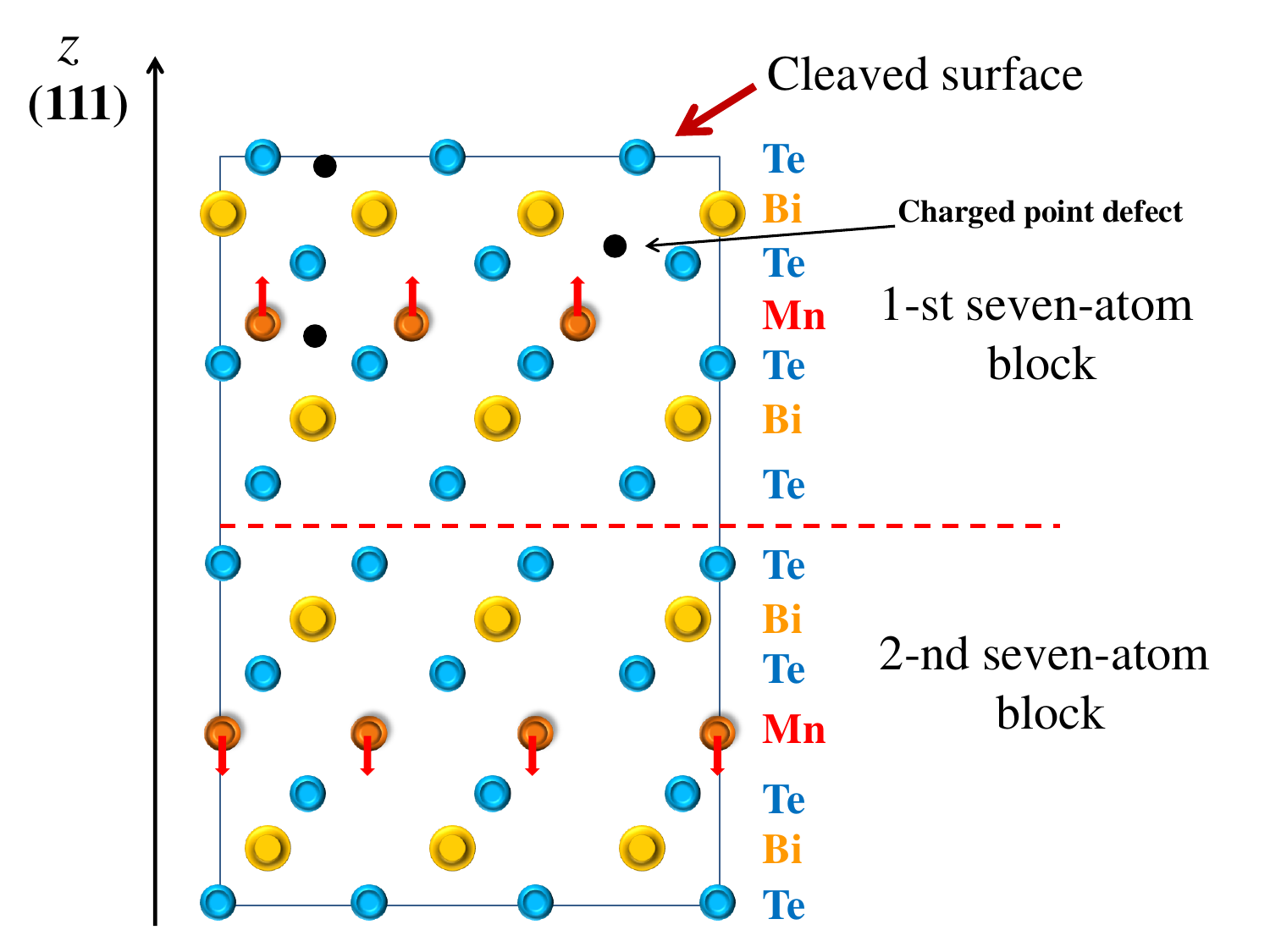}
\caption{Schematic structure of the antiferromagnetic topological insulator MnBi$_2$Te$_4$. The red up and down arrows on the circles represent spin up and spin down AFM magnetization, the black dots represents charged impurities at the surface layer. Surface orientation corresponds to the (111) crystallographic plane.}
\label{fig1}
\end{figure}

The magnetic order in the TIs can be introduced either by doping a non-magnetic TI with magnetic atoms or by synthesis of the stoichiometric TI with magnetic ions in its crystal structure. The latter approach looks more promising since it allows to obtain homogeneous samples. The first synthesized intrinsic MTI was MnBi$_2$Te$_4$~\cite{otrokov2019prediction,zhang2019topological,gong2019experimental,PhysRevResearch.1.012011,aliev2019novel,hao2019gapless,chen2019topological,swatek2020gapless,Liang2022,Qiu2023,Gao2023,Wang2023,Chen2023,Cao2023}, which is currently being intensively studied. This material has a layered Van-der-Waals structure. Each seven-atom block or layer of MnBi$_2$Te$_4$ can be schematically written as Te-Bi-Te-Mn-Te-Bi-Te (see Fig.~\ref{fig1}). Magnetic ions of Mn are ferromagnetically ordered within the layer, and the layers are ordered antiferromagnetically (AFM). The Neel temperature for MnBi$_2$Te$_4$ is 25~K~\cite{otrokov2019prediction,PhysRevResearch.1.012011}, which is the largest among existing MTIs. 

In the case of AFM MTIs, it is an open question whether the spectrum of the surface states has a gap or not. Density functional theory (DFT) calculations predict the gap spectrum for the surface states~\cite{otrokov2019prediction,PhysRevLett.122.206401,Li2019}. The ARPES measurements give a wide range of the gap from 0~\cite{swatek2020gapless} to 100~meV~\cite{zeugner2019chemical}. Such a scattering of results can be attributed to differences in sample quality. Other explanations include the surface electrostatic field~\cite{Shikin2021}, increasing the interlayer distance between the van der Waals block near the surface~\cite{shikin2020nature}, and native point defects~\cite{garnica2022native}. Thus, the influences of material parameters and defects on the electronic gap in the spectrum of MTIs are important issues.
To address these issues, the {\it ab initio} calculations of the energy spectrum of MnBi$_2$Te$_4$ were performed~\cite{shikin2020nature} as well as the analytical analysis based on the effective Hamiltonian of the system in the ${\mathbf{k\cdot p}}$ approximation~\cite{sun2020analytical,PhysRevLett.122.206401}. The authors of Ref.~\cite{sun2020analytical} obtained an analytical dependence of the surface energy gap on the bulk properties of the MTI. They conclude that a possible reason for the scattering of the experimental data could be attributed to the idea that in real materials the intralayer ferromagnetic order is smaller and more localized than implied by the theoretical analysis. In the seminal paper Ref.~\cite{Mong2010}, based on a toy model for the topological insulator, the authors predicted that the spectrum of surface states in the AFM phase is gapped. Using a different model, the authors obtained that the surface states are detached from the bulk states, implying a rich physics of the surface states in the MTIs in the AFM state depending on the underlying model. In Ref.~\cite{Walko2022} it was shown that presence of the gap depends on the termination of the surface. All of the above allows us to assert that at present there is no generally accepted understanding of the nature of the gap in the spectrum of surface states in the AFM MTIs.

In Ref.~\cite{Ma2020} it was observed that the surface doping opens a gap in the spectrum of the surface states. The DFT calculations performed in Refs.~\cite{shikin2020nature,shikin2022electronic} predict that the charged impurities at the surface control the surface band gap in MnBi$_2$Te$_4$. Therefore, investigating of the influence of the surface doping on the gap in the surface states is an important task.

In this paper we study the surface states of the AFM MTI. We start with the linearised in momentum MTI Hamiltonian, in which we explicitly consider a spatial variation of the magnetisation across the layers of the material. We then calculate an effective Hamiltonian for the surface states. The Hamiltonian obtained differs from one derived in Ref.~\cite{PhysRevLett.122.206401} since it has an extended basis that includes the indices of the AFM-ordered layers. The effective Hamiltonian for the surface states has a typical Dirac form and the spectrum is gapless. This gapless spectrum is protected by the combined time-reversal symmetry which consists of time-reversal symmetry and half-translation symmetry~\cite{PhysRevLett.122.206401}. However, if we break such a symmetry by perturbing the surface, the gap at the Dirac point appears. We show that the shift of the chemical potential of the surface layer opens a gap in the spectrum at the Dirac point due to the AFM ordering. This gap is robust to disorder, that is, even strong disorder does not suppress the gap to zero. 
We extend the bulk MTI Hamiltonian to include quadratic out-of-plane momentum terms. We find that such terms violate the combined time-reversal symmetry that allows for a finite gap without surface perturbation. We calculate the spectrum of such a Hamiltonian in a tight-binding approximation and observe a finite gap for the surface states.
We discuss the consistency of our results with the experiment and DFT calculations.     

\section{Model}\label{Model}

We are interested in the electron spectrum of the AFM MTI of the type MnBi$_2$Te$_4$ near the $\Gamma$-point. Following Ref.~\cite{PhysRevLett.122.206401}, we start with a low-energy Hamiltonian in the form
\begin{equation}\label{W}
H_0=-\mu+m\sigma_z+v(k_xs_x+k_ys_y)\sigma_x+v_zk_zs_z\sigma_x,
\end{equation}
where we neglect terms of order $\mathbf{k}^2$ and higher. Here $k_i$ are the components of the momentum $\mathbf{k}$, $v$ and $v_z$ are the in-plane and transverse components of the Fermi velocity, respectively. The Pauli matrices $s_i$ act in the spin space $(\uparrow, \downarrow)$ and the Pauli matrices $\sigma_i$ act in the space of the low-energy orbitals $\left(|P1^+_z\rangle,|P2^-_z\rangle\right)$, where the superscripts $\pm$ stand for the parity of the corresponding states~\cite{PhysRevLett.122.206401}. In the absence of spin-orbit coupling, the states $|P1^+_z\!\uparrow\!(\downarrow)\rangle$ are associated with two Bi orbitals, while $|P2^-_z\!\uparrow\!(\downarrow)\rangle$ are associated with two Te orbitals. We will also use more descriptive notations, Bi$_{\uparrow(\downarrow)}$ and Te$_{\uparrow(\downarrow)}$, for the low-lying states considered. The spectrum of the Hamiltonian~\eqref{W} at $k_z=0$ is $E_0=-\mu\pm\sqrt{m^2+v^2k^2}$, and the physical meaning of $m$ is a gap in the spectrum in the bulk. The Hamiltonian~\eqref{W} is equivalent to the Hamiltonian for an ordinary TI~\cite{Liu2010} up to the rotation of the basis.   

To describe the magnetic ordering, we should add to $H_0$ the corresponding magnetic terms. To take into account the spacial structure of the AFM ordering explicitly, one has to consider two seven-atom blocks with Mn atoms having opposite directions of the magnetic moment, which requires extension of the Gilbert space of Hamiltonian~\eqref{W} from 4D to 8D. However, the authors of Refs.~\cite{sun2020analytical,PhysRevLett.122.206401} made a projection of the space that neglected this feature. Such an approach allows them to reduce the space of the Hamiltonian from 8D to 4D.

In order to restore information on the spatial structure of the AFM state, we introduce an additional Gilbert space $t$ that takes into account pairs of the magnetic layers with opposite polarization of Mn atoms. In other words, we add additional degree of freedom that can be associated with the even and odd layers in the same way as spin up and spin down regions for the spin degree of freedom. Our approach in this part is similar to the approach from Ref.~\cite{Mong2010}. First, we transform the kinetic energy term in the Hamiltonian $H_0$. This transformation is $v_zk_zs_z\sigma_x\rightarrow v_zk_zs_z\sigma_xt_x$, where $t_x$ is the Pauli matrix that acts in the $t$ space, since only the nearest-neighbour hoping between the Bi and Te orbitals is allowed. In the absence of the AFM order we have half-translational symmetry between even and odd layers $[\hat{H},\hat{S}]=0$, where operator of the half-translation is $\hat{S}=t_x$. In general, the AFM ordering results, first, in the finite magnetization $M_z$ of each seven-atom layer and, second, in a spin imbalance between Bi and Te orbitals $A_z$. Such terms naturally break the symmetry between even and odd layers. The corresponding term in the Hamiltonian reads 

\begin{equation}\label{hm1}
 H_m=M_zs_zt_z+A_zs_z\sigma_zt_z.    
\end{equation}
This term includes, in particular, the spatial dependence of the magnetization: for even and odd layers we have magnetization with the opposite directions. 

Thus, the Hamiltonian of the AFM MTI in the extended space is
\begin{eqnarray}\label{Ham}
H=&-&\mu+m\sigma_z+v(k_xs_x+k_ys_y)\sigma_x\nonumber \\
&+&v_zk_zs_z\sigma_xt_x +M_zs_zt_z+A_zs_z\sigma_zt_z.
\end{eqnarray}
The Pauli matrices $t_i$,$i=x,y,z$ act in the space of even/odd layers.
As we can see, the magnetic terms break the time-reversal symmetry $\hat{T} H({\bf{k}}) \hat{T}^{-1} \neq H(-\bf{k})$, where $\hat{T}=is_yK$ and $K$ is the complex conjugation. The translation symmetry between layers $1$ and $2$ is also broken by the magnetic terms: $\hat{S} H({\bf{k}}) \hat{S}^{-1}\neq H(\bf{k})$, where operator of the half-translation symmetry is $\hat{S}=t_x$. However, a combined time-reversal symmetry $\hat{T}_a H({\bf{k}}) \hat{T}_a^{-1} = H(-\bf{k})$ is preserved, where $\hat{T}_a=\hat{S}\hat{T}=is_yt_xK$ and $\hat{T}_a^2=-1$. This symmetry ensures Kramer's degeneracy of the spectrum.

\section{Surface states}

We assume that the sample occupies the space $z>0$ and its surface lies in the $(x,y)$ plane at $z=0$ that corresponds to the crystallographic plane $(111)$ of the natural cleavage. We replace $k_z$ by the operator $-i\partial_z$ in the Hamiltonian ($\hbar=1$) and put $k_x,k_y=0$ in the main approximation in $\mathbf{k}$. As a result, the surface states $\Psi(z)$ obey the equation
\begin{eqnarray}\label{sys}
\!\!\!\!\!\!\!\!\!\left[m\sigma_z\!+\!s_z(M_zt_z\!+\!A_z\sigma_zt_z\!-\!iv_z\sigma_xt_x\partial_z)\!-\!\mu\right]\Psi=E\Psi,
\end{eqnarray}
where $E$ is the energy. We choose $E=0$ since we are interested in the states near the Dirac point. We seek the solution to the problem in the form of an eighth-component spinor
\begin{equation}\label{Psi}
\Psi = (\textrm{Bi}_{\uparrow1},\textrm{Bi}_{\downarrow1},\textrm{Te}_{\uparrow1},\textrm{Te}_{\downarrow1},
\textrm{Bi}_{\uparrow2},\textrm{Bi}_{\downarrow2},\textrm{Te}_{\uparrow2},\textrm{Te}_{\downarrow2}).
\end{equation}
Each component of the spinor is characterised by three quantum numbers: the orbital index $\sigma=\textrm{Bi,\,Te}$, the spin projection $s=\uparrow,\downarrow$ and the magnetic layer number $t=1,2$. 
In addition, the normalisation condition $\int_0^{+\infty}|\Psi(z)|^2 dz=1$ should be satisfied.

The boundary problem should be accompanied by a proper boundary condition for the wave function. We start with the discussion of non-magnetic TI. A uniform boundary condition $\Psi(0)=0$ for non-magnetic TI was proposed in Ref.~\cite{Liu2010}. However, in the linear approximation in $\partial_z$ such a problem has only a trivial solution. In Refs.~\cite{Fu2010,PhysRevLett.108.107005} the authors found that the van der Waals system Bi$_2$Se$_3$ is naturally cleaved between two quintuple layer blocks. This allows the formulation of appropriate boundary conditions. Obviously a similar situation is realised for MnBi$_2$Te$_4$, but now we have the seven layer unit and two AFM ordered blocks. Following the approach of Refs.~\cite{Fu2010,PhysRevLett.108.107005}, we express the boundary conditions in our matrix notation in the form 
\begin{equation}\label{bound}
(1+\sigma_x)(1+t_z)\Psi(0)=0,\quad \Psi(+\infty)=0.
\end{equation} 
Physically, this condition implies that only one magnetic layer and only one orbital on a proper basis reach the surface. 
Such boundary condition preserves time-reversal and $C_3$ rotational symmetries. Linear equations~\eqref{sys} along with the boundary conditions Eq.~\eqref{bound} and normalization conditions form a complete system of equations that allows us to calculate the spinor $\Psi$ for the surface states. We obtain two linearly independent solutions. It is convenient to introduce an orthonormal basis $(\Psi_1,\Psi_2)$, and any solution $\Psi$ is a linear combination of $\Psi_1$ and $\Psi_2$. We choose $\Psi_1$ and $\Psi_2$ in the form
\begin{align}\label{basis}
&\Psi_1=\textrm{Bi}^{(0)}_{\uparrow1}\left(e^{\lambda_1z},0,-e^{\lambda_2z},0,-i\kappa_1e^{\lambda_2z},0,i\nu_2e^{\lambda_1z},0\right),\\
\nonumber
&\Psi_2=\textrm{Bi}^{(0)}_{\downarrow1}\left(0,e^{\lambda_2z},0,-e^{\lambda_1z},0,-i\nu_1e^{\lambda_1z},0,i\kappa_2e^{\lambda_2z}\right),
\end{align}

where 
\begin{align}
&\textrm{Bi}^{(0)}_{\uparrow 1}\!=\!\left[\frac{M_z+m}{|\lambda_1|(M_z\!+\!m\!-\!A_z\!+\!\mu)}\!\!+\!\!\frac{M_z-m}{|\lambda_2|(M_z\!-\!m\!+\!A_z\!+\!\mu)}\!\right]^{\!-\!\frac 12}\!,\\
&\textrm{Bi}^{(0)}_{\downarrow1}=\!\left[\frac{M_z+m}{|\lambda_1|(M_z\!+\!m\!+\!A_z\!-\!\mu)}\!\!+\!\!\frac{M_z-m}{|\lambda_2|(M_z\!-\!m\!-\!A_z\!-\!\mu)}\right]^{-\frac 12},\nonumber\\ 
&\kappa_{1,2}= \sqrt{\frac{M_z-m\mp A_z\mp \mu}{M_z-m\pm A_z\pm\mu}},\quad
\nu_{1,2}= \sqrt{\frac{M_z+m\mp A_z\mp \mu}{M_z+m\pm A_z\pm\mu}}\nonumber\\
&\lambda_{1,2}=-\frac{1}{v_z}\sqrt{(M_z\pm m)^2-(A_z\mp \mu)^2}\nonumber.
\end{align}
The surface states exist if Re$\lambda_{i}<0$ for both $i=1,2$. This constraint imposes conditions on the values of the parameters under which the surface states can exist. We checked that the current does not flow through the plane.

\section{Effective Hamiltonian for the surface states}\label{sec::ham}

To derive an effective surface Hamiltonian $H_{\textrm{s}}$ we make a projection of the Hamiltonian~\eqref{Ham} on the basis vectors Eqs.~(\ref{basis}), $H_{\textrm{s}}=\langle\Psi_i|H|\Psi_j\rangle$. After integration over $z$, and considering $v(k_xs_x+k_ys_y)\sigma_x$ as a perturbation we obtain a typical Dirac-like Hamiltonian for the surface states
\begin{eqnarray}\label{H_s}
H_{\textrm{s}}&=&\tilde{v}(k_x\hat{s}_x+k_y\hat{s}_y),\\
\tilde{v}&=&v\frac{\textrm{Bi}^{(0)}_{\uparrow1}\textrm{Bi}^{(0)}_{\downarrow1}(|\lambda_1|+|\lambda_2|)}{|\lambda_1\lambda_2|},  \nonumber 
\end{eqnarray}
where $\hat{s}$ are the Pauli matrices in the space of the vectors $\Psi_1$ and $\Psi_2$. The Hamiltonian has a linear gapless spectrum $E=\pm\tilde{v}\sqrt{k_x^2+k_y^2}$. If we assume that the bulk gap is large, $m\gg |M_z|,|A_z|$, we derive $\tilde{v} = v(1-\mu^2/m^2) + O(M_z^2)+O(A_z^2)$. 

The wave function $\Psi_1$ corresponds to the orbitals with the real spin projection $s_z=\uparrow$, while $\Psi_2$ corresponds to $s_z=\downarrow$, see Eqs.~\eqref{basis}. In addition, in our orthonormalized basis $\hat{s}_\alpha\propto\langle\Psi_i|s_\alpha|\Psi_j\rangle$. Therefore, we can consider the Pauli matrices in the space of the surface states $\hat{s}$ as real-spin operators. Note that a similar result for the surface states was obtained in Ref.~\cite{Liu2010}.  

We see that the AFM ordering affects only quantitatively the surface states, as compared to the non-magnetic case $M_z=A_z=0$~\cite{sun2020analytical,PhysRevLett.122.206401}. This result is not surprising: in both non-magnetic and AFM cases, the system has the same symmetries except the time-reversal symmetry ${\hat T}$. However, the AFM TI has emergent time-reversal-like symmetry ${\hat T}_a$ in the extended space, as it was stated in Section~\ref{Model}. Therefore, we need to break symmetry between layers 1 and 2 with different polarizations of Mn atoms to observe a qualitatively new result. The simplest way to break this symmetry is to introduce a difference in the chemical potential between layers 1 and 2 due to surface doping. For simplicity, we assume that the chemical potential in the bulk and in the second seven-atom block is the same, while it is different in the single surface block. We introduce the operator of the surface chemical potential as
\begin{equation}\label{eq::mus}
\hat{\mu}_{\textrm{s}}=-\mu_{\textrm{s}}\frac{\hat{1}+t_z}{2},   
\end{equation}
where factor $(1+t_z)/2$ selects single seven-atom block 1 as the surface termination and $\mu_s$ is a difference between chemical potentials in the layers 1 and 2. Now we calculate matrix elements $\hat{\mu}_{ij}=\langle\Psi_i|\hat{\mu}_{\textrm{s}}|\Psi_j\rangle$ that gives us
\begin{eqnarray}\label{surf_doping}
\hat{\mu} &=& -\tilde{\mu}_{\textrm{s}}(1 + \delta\hat{s}_z), \\
\nonumber
\tilde{\mu}_{\textrm{s}}&=&\mu_{\textrm{s}}\frac{\left(\textrm{Bi}_{\uparrow1}^{(0)2}+\textrm{Bi}_{\downarrow1}^{(0)2}\right)(|\lambda_1|\!+\!|\lambda_2|)}{2\sqrt{2}|\lambda_1\lambda_2|}, \\
\nonumber
\delta &=& \frac{\textrm{Bi}_{\uparrow1}^{(0)2}-\textrm{Bi}_{\downarrow1}^{(0)2}}{\textrm{Bi}_{\uparrow1}^{(0)2}+\textrm{Bi}_{\downarrow1}^{(0)2}}.
\end{eqnarray}

When $m\gg |M_z|,|A_z|$ we get $\delta= M_z\mu/m^2(1+\mu^2/2m^2)+A_z/m(1/2+\mu^2/m^2)$. 
We can see that the shift of the surface chemical potential brings the term $\propto \hat{s}_z $ that opens a gap in the spectrum of the surface states of the MTI. The effective Hamiltonian~\eqref{H_s} with the surface perturbation now reads as
\begin{equation}\label{ham}
H = \tilde{v}(k_x\hat{s}_x+k_y\hat{s}_y) - {\tilde\mu}_{\textrm{s}}(1+\delta\hat{s}_z).  
\end{equation}
The spectrum of this Hamiltonian is
\begin{equation}\label{E(k)}
E_{\pm} = -{\tilde\mu}_{\textrm{s}}\pm \sqrt{\tilde{v}^2k^2+{\tilde\mu}_{\textrm{s}}^2\delta^2}.
\end{equation}
The finite $\delta$ induces a gap between $E_{+}$ and $E_{-}$. This term arises due to the finite charge imbalance between the top layer and the next layer, given by the $t_z$ term in Eq.~\ref{eq::mus}. 

\begin{figure}[ht]
\includegraphics[width=1.0\linewidth]{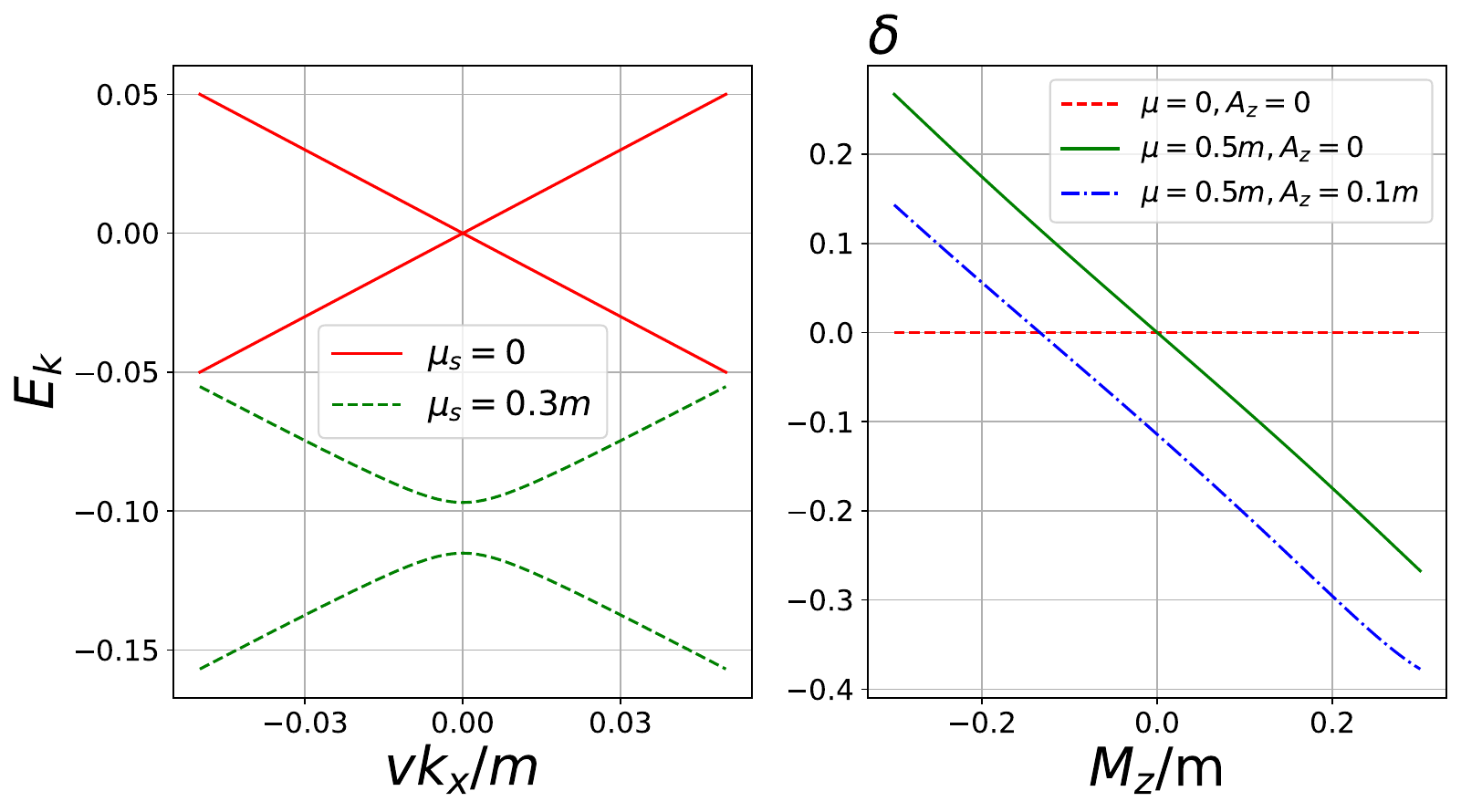}
\caption{Left panel: energy spectrum as a function of the momentum $k_x$ and $k_y=0$. We take $\mu=0.5m$, $M_z=0.1m$, $A_z=0$. Spectrum have a gap if $\mu_s \neq 0$. Right panel: parameter $\delta$ as a function of the AFM magnetization $M_z$.}
\label{d}
\end{figure}

We plot in Fig.~\ref{d} (right panel) the parameter $\delta$ that controls the gap in the surface spectrum as a function of the AFM magnetization $M_z$ for different values of the bulk chemical potential $\mu$ and $A_z$. We see that $\delta\propto M_z$, $\mu$ controls the slope of this line, and $A_z$ shifts $\delta(M_z)$ from the coordinate origin. The larger are $\mu$ and $M_z$, the larger is the surface electron gap. The spectrum $E(\mathbf{k})$, Eq.~\eqref{E(k)}, is shown in Fig.~\ref{d} (left panel) for different values of ${\tilde\mu}_{\textrm{s}}$. When we ignore the AFM order, $M_z=A_z=0$, the gap vanishes since $\delta=0$ in this case.

\section{Tight-binding model with the quadratic terms}
\label{sec::tb}

\begin{figure}[ht]
\includegraphics[width=1.0\linewidth]{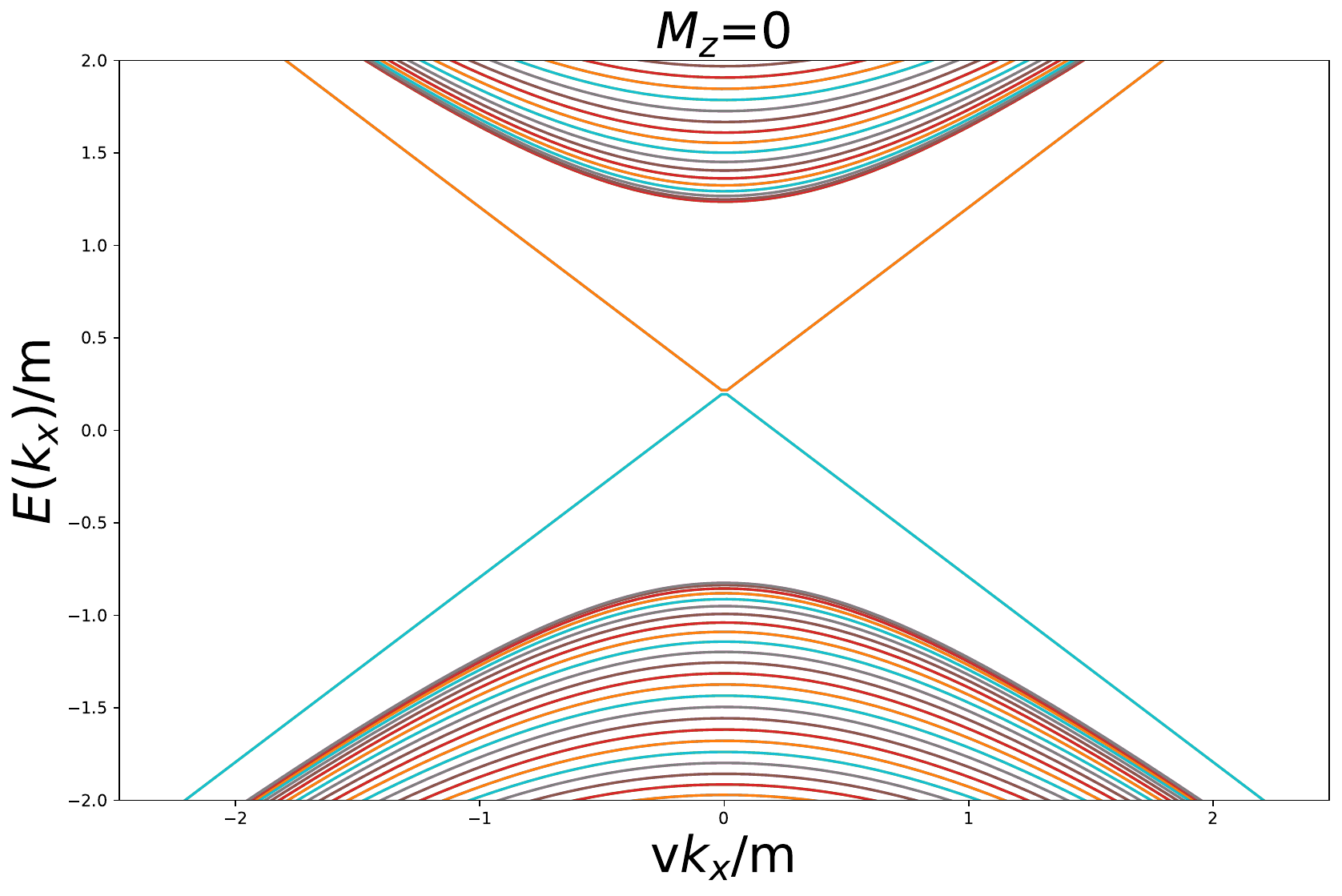}
\includegraphics[width=1.0\linewidth]{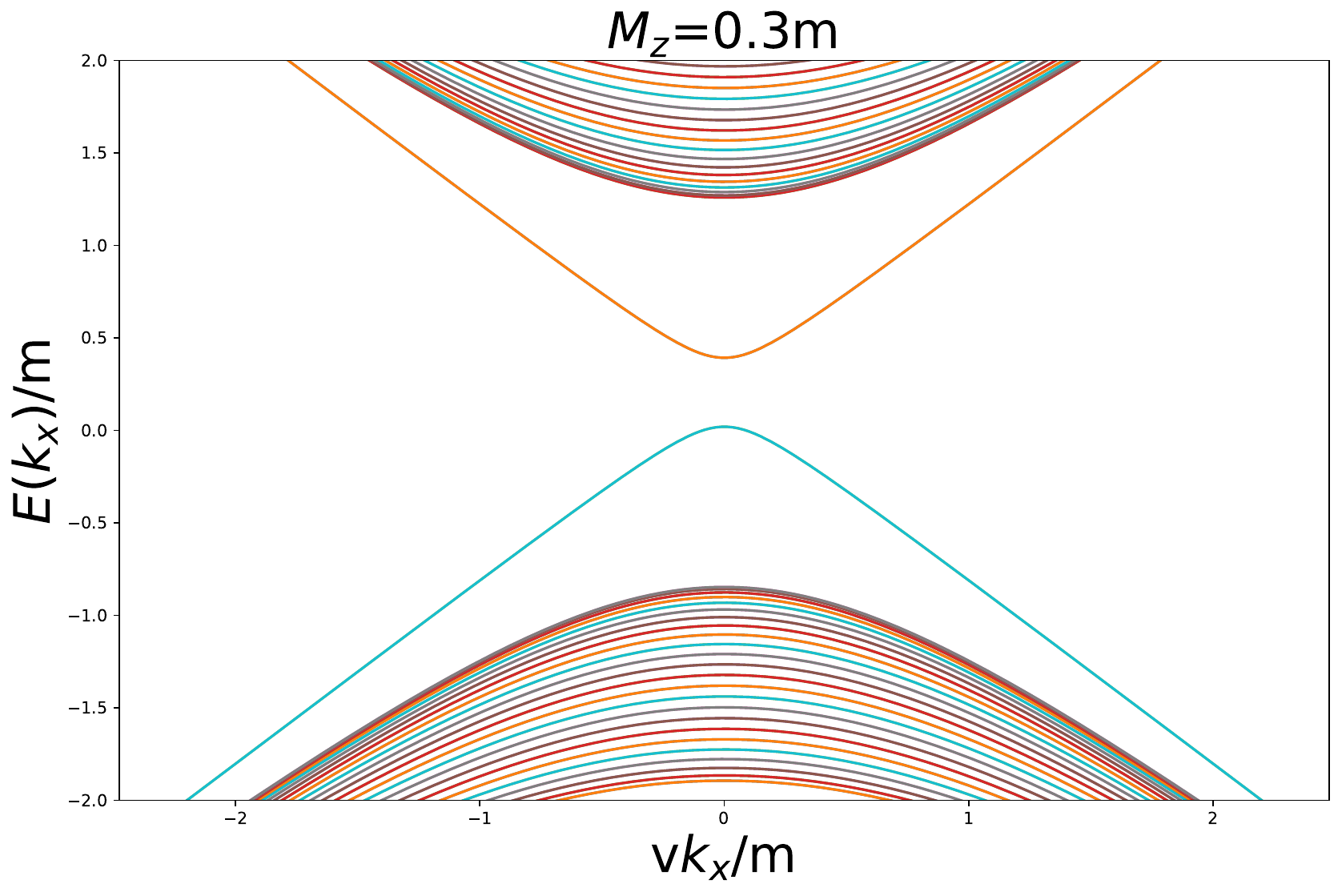}
\caption{The spectrum $E(k_x)$ of the tight-binding Hamiltonian for $k_y=0$ as a function of the in-plane momentum $vk_x/|m|$. Upper figure corresponds to $M_z=A_z=0$. Lower figure corresponds to $M_z=0.3m$,$A_z=0$. Following Ref.~\cite{PhysRevLett.122.206401} we take $m=-0.12eV$, $\mu=-0.2m$, $v=3.2 eV\cdot \AA$,$v_z=2.7 eV\cdot \AA$, $C_1=2.7 eV\cdot \AA^2$, $C_2=17 eV\cdot \AA^2$,  $M_1=12 eV\cdot \AA^2$, $M_2=9.4 eV\cdot \AA^2$. Also, we set $\mu=-0.2m$ and lattice constant $a=15\AA$ which is smaller than actual lattice constant $a\sim40\AA$~\cite{Yan2019} in order to get gapless surface states without magnetization. The calculations are performed for 200 layers.}
\label{tbs}
\end{figure}

In this section we numerically study the energy spectrum of the tight-binding model based on the Hamiltonian~\eqref{W} with the AFM terms given by Eq.~\eqref{hm1}. We also extend the model to include the quadratic terms in the momentum. Following Ref.~\cite{PhysRevLett.122.206401} we replace $\mu\rightarrow \mu+C_1k_z^2+C_2(k_x^2+k_y^2)$ and $m\rightarrow m+M_1k_z^2+M_2(k_x^2+k_y^2)$. 
The quadratic terms in $k_z$ double the order of the differential equations~\eqref{sys}. In this case the boundary conditions given by Eq.~\eqref{bound} are no longer sufficient. Instead we use the following boundary conditions~\cite{Liu2010,PhysRevLett.122.206401}
\begin{equation}\label{bound2}
(1+t_z)\Psi(0)=0, \qquad \Psi(+\infty)=0, 
\end{equation} 
assuming that the wave function $\Psi$ vanishes at the surface and in the bulk. We map the low-energy Hamiltonian to the tight-binding model and numerically calculate the electron spectrum of the surface states. The result is shown in Fig.~\ref{tbs}. We see that the finite AFM order opens a gap in the spectrum of the surface states. This result is in agreement with the tight-binding calculations made in~\cite{shikin2020nature,shikin2022electronic}.

To get an idea of the appearance of the gap when we consider quadratic momentum, we explain in detail in Appendix A how we extend the model basis to include the degrees of freedom associated with the layers having opposite magnetic momenta. We find that the low energy Hamiltonian is written now as
\begin{equation*}
\!\!\!\!\!\!\!\! H_2 \!=\!H+ C_1k_z^2t_x+C_1k_z t_y/a + M_1k_z^2\sigma_zt_x+M_1k_z\sigma_zt_y/a,    
\end{equation*}
where $H$ is given by Eq.~\eqref{Ham}.

We see that we have terms in both $t_x$ and $t_y$. If we include only the linear term in $k_z$ in the Hamiltonian~\eqref{general} and exclude the quadratic corrections to the Hamiltonian $C_1=M_1=0$, we have only the $t_x$ part. This case is realised in the Hamiltonian~\eqref{Ham}. The quadratic in $k_z$ terms give rise to $t_y$ terms in the Hamiltonian, which preserve the time-reversal symmetry $\hat{T}$ but violate the half-translation symmetry $[\hat S,t_y]=[t_x,t_y] \neq 0$. This violates the extended time reversal symmetry $\hat{T}_a=\hat{T}\hat{S}$, which leads to the finite gap observed in the calculations.

\section{Discussion}

In this work we investigate the effects of the antiferromagnetic ordering on the properties of the surface states in the topological insulator. We get that the change in chemical potential at the surface layer opens a gap in the spectrum away from the Fermi energy due to the finite AFM order. Such a gap is controlled by the value of the surface doping. The gap is robust against disorder. These results are consistent with the experiments that demonstrate opening the surface gap by the surface doping~\cite{Ma2020}. We should note that in a real samples the effect of the surface defects can be more significant than just electron density disorder and can lead to a quite complex picture~\cite{Tan2023}.

In non-magnetic topological insulators the gapless Dirac spectrum of the surface states is protected by the time-reversal symmetry $\hat{T}$. 
Antiferromagnetic order parameter breaks both time-reversal symmetry and symmetry between even and odd layers. Instead, a combined time-reversal-like symmetry $\hat{T}_{a}=\hat{t}_x\hat{T}$ is preserved. The latter symmetry includes translation between spin-up and spin-down AFM layers~\cite{PhysRevLett.122.206401}. Such a symmetry protects gapless spectrum of the surface states. In order to open a gap in the spectrum this symmetry has to be broken. Since the combined symmetry consist of time-reversal symmetry and translational symmetry we can break any of them. We can violate time-reversal symmetry by the out-of-plane magnetization and open the gap in the similar manner to the topological insulators without the AFM order. The other way is to break symmetry between spin-up and spin-down layers. Unlike real time-reversal symmetry, this combined symmetry can be broken by a non-magnetic perturbation (see, e.g., Eq.(9) to check that $[\hat{T}_{a},\hat{\mu}] \neq 0$). If the perturbation breaks symmetry between spin-up and spin-down regions, then, a gap in the spectrum of the surface states can be formed, see Fig.~\ref{d}. Such mechanism opens a gap in the spectrum only if AFM order is present in the system.

Experimentally, surface disorder is caused by natural exposure of the surface to the atmosphere~\cite{Brahlek2011} or by external doping of the surface~\cite{Ma2020}. Recent DFT calculations of the properties of MnBi$_2$Te$_4$ show that doping the first surface layer with Ge increases the gap in the surface states~\cite{Estyunina2023}. Also, DFT calculations performed in Ref.~\cite{shikin2022electronic} show that the surface potential opens a gap in the surface states in the AFM MTI. Experimentally, Sb doping of the surface of MnBi$_2$Te$_4$ opens a gap in the spectrum of surface states~\cite{Ma2020}.
We have considered the shift of the surface chemical potential due to an effective surface doping, neglecting the possible disorder of such a perturbation. Therefore, an important question is the stability of the surface gap against disorder. Note that in the case of a magnetisation-induced gap in the surface states of non-magnetic TI, the strong disorder suppresses the gap~\cite{Akzyanov2023}. In Appendix. B we have shown that the gap in the spectrum of surface states given by Eq.~\eqref{ham} remains finite even for large values of disorder.  

The results of the DFT calculations presented in Refs.~\cite{otrokov2019prediction,PhysRevLett.122.206401,Li2019} predict a gap in the spectrum of surface states of the AFM MTI without any surface disorder. These results are consistent with our tight-binding calculations in Sec.~\ref{sec::tb}. The reason for the gap opening is the violation of the combined time-reversal symmetry, which is possible due to different boundary conditions. Further studies are needed to determine which model better describes the experimental results.

\section{Conclusions}
In this paper we have derived the Hamiltonian of the bulk states of the magnetic topological insulator which includes the antiferromagnetic order parameter. For this Hamiltonian, we obtain the effective Hamiltonian of the surface states, which has the same form as for the nonmagnetic topological insulator. The surface doping opens a gap in the spectrum of surface states due to the finite antiferromagnetic order. We perform tight-binding calculations for different boundary conditions and obtain the gapped spectrum for the surface states.

\section*{Declaration of competing interest}
The authors declare that they have no known competing financial interests or personal relationships that could have appeared to influence the work reported in this paper.

\section*{Acknowledgments}

This work is supported by Russian Science Foundation (project № 22-72-10074). 
\bibliographystyle{elsarticle-num}
\bibliography{BoundS}
\appendix

\setcounter{equation}{0}
\renewcommand{\theequation}{A.\arabic{equation}}
\section{Derivation of the low energy Hamiltonian in a extended space with quadratic terms}
\label{sec:b}
We start with the low energy Hamiltonian, which is presented in a general form
\begin{equation}\label{general}
H_g=H_0+\alpha_0 k_z+\beta_0 k_z^2,
\end{equation}
where $H_0$, $\alpha$, $\beta$ are some hermitian matrices.
Here we do a substitution $k_z a \rightarrow \sin k_za$ and $(k_z a)^2 \rightarrow 2(1-\cos k_za)$, where $a$ is the lattice constant, and get
\begin{equation}
H_g=H+\alpha \sin k_za+\beta \cos k_za.
\end{equation}
Here $H=H_0+2\beta_0/a^2$, $\alpha = \alpha_0/a$, $\beta=-2\beta_0/a^2$.
This Hamiltonian corresponds to the following tight-binding model in real space
\begin{align}
H_{tb}=h_0+h_1+h_2 \\
h_0 = \sum\limits_na_{n}^{\dagger}Ha_n+h.c. \\
h_1 = \sum\limits_n(a_{n+1}^{\dagger} \alpha a_n - h.c)/2i,\\
h_2 = \sum\limits_n(a_{n+1}^{\dagger} \beta a_n + h.c)/2, 
\end{align}
where $a_n$ is the annihilation operator of the electron in the layer $n$.
We introduce new annihilation operators for odd and even layers $c_n = a_{2n+1}$, $d_n=a_{2n}$. We focus only on the terms $h_1$ and $h_2$, since $h_0$ is trivially transformed:
\begin{align}
h_1 = \sum\limits_n(c_{n+1}^{\dagger} \alpha d_n +d_{n}^{\dagger} \alpha c_n - h.c)/2i,\\
h_2 = \sum\limits_n(c_{n+1}^{\dagger} \beta d_n +d_{n}^{\dagger} \beta c_n + h.c)/2.
\end{align}

We perform a Fourier transform on the operators $c_k = \sum c_n \exp{(-2ik_zna)}$ and $d_k = \sum d_n \exp{(-2ik_zna)}$, taking into account that the length of the elementary lattice is doubled when we introduce even and odd layers. 
After that we have  
\begin{align}
h_1 = (c_k^{\dagger}\alpha d_k e^{2ik_za}+d_k^{\dagger}\alpha c_k - h.c)/2i,\\
h_2 = (c_k^{\dagger}\beta d_k e^{2ik_za}+d_k^{\dagger}\beta c_k + h.c)/2.
\end{align}
We can rewrite the result in the matrix form  
\begin{equation}
h_{1,2} = \begin{pmatrix}
c^{\dagger}\,d^{\dagger}
\end{pmatrix}
H_{1,2}
\begin{pmatrix}
c\\d
\end{pmatrix}     
\end{equation}
where 
\!\!\!\begin{equation}
H_1\!=\!\frac{\alpha}{2}\!
\begin{pmatrix}
0&\sin 2k_za \!-\!i(1\!-\!\cos 2k_za)\\
\sin 2k_za \!+\!i(1\!-\!\cos 2k_za)&0
\end{pmatrix}
\end{equation}
and
\begin{equation}
\!\!\!H_2\!=\!\frac{\beta}{2}\!
\begin{pmatrix}
0&1+\cos 2k_za+i\sin 2k_za\\
1+\cos 2k_za-i\sin 2k_za&0
\end{pmatrix}
\end{equation}
We introduce Pauli matrices $t_{i}$, $i=0,x,y,z$ acting in the odd/even layer space $(c,d)$. We extend the sine and cosine terms up to $O(k_z^3)$ and obtain a low energy Hamiltonian in the extended space
\begin{equation}
H_{ge} = H_0+\alpha_0k_z (t_x+ t_yk_za)+2\beta_0k_z(t_x k_z+ t_y/a).
\end{equation}

\section{Effects of the disorder}
\label{sec:b}
We consider a short-range disorder near the sample surface of the AFM MTI produced by randomly distributed charged point defects, see Fig.~\ref{fig1}. Such a perturbation can be more intensive in the surface layer 1 since it is naturally least protected from external influences. We denote 2D density of the point defects as $n$ and the local impurity potential at the position $\mathbf{r}=\mathbf{R}_j$ as $u_j$. By analogy with Eq.~\eqref{surf_doping}, the operator of the disorder potential in the basis of the surface states has a form $\hat{U}=\sum_j \hat{U}_j$, where $\hat{U}_j(\mathbf{r})=(\hat{1}+\delta t_z)u_j\delta(\mathbf{r}-\mathbf{R}_j)/2$ and $\delta(\mathbf{r})$ is the delta function. We assume that the disorder is Gaussian, that is, $\langle \hat{U}_j(\mathbf{r})\rangle=0$ and $\langle \hat{U}_i(\mathbf{r}_1) \hat{U}_j(\mathbf{r}_2) \rangle=n u_0^2 \delta (\mathbf{r}_1-\mathbf{r}_2)\delta_{ij}$, where $\langle ...\rangle$ means the spatial average, $u_0^2=\langle u_i^2\rangle$, and $\delta_{ij}$ is the Kronecker symbol. 

We assume that the disorder is weak, that is, $j=nu_0^2/(2\pi \tilde{v}^2)<1$ and, following a standard procedure, calculate the self-energy $\hat{\Sigma}$ in the Born approximation. As a result, we obtain in the  n-th order a recursive Born series
\begin{equation}\label{born_eq}
\begin{cases} 
\hat{\Sigma}^{(n+1)}= \sum\limits_{k,i} \langle \hat{U}_i G(\hat{\Sigma}^{(n)}) \hat{U}_i \rangle, \\
G^{-1}(\hat{\Sigma}^{(n)})=-H-\hat{\Sigma}^{(n)}.\end{cases}
\end{equation}
If this procedure converges and $\hat{\Sigma}^{(n)}\rightarrow \hat{\Sigma}$ for $n \rightarrow +\infty$, then, the sum of the series can be represented as a self-consistent Born approximation (SCBA) solution: $\hat{\Sigma}= \sum_{i} \langle \hat{U}_i G(\hat{\Sigma}) \hat{U}_i \rangle$.

\begin{figure}[ht]
\includegraphics[width=1.0\linewidth]{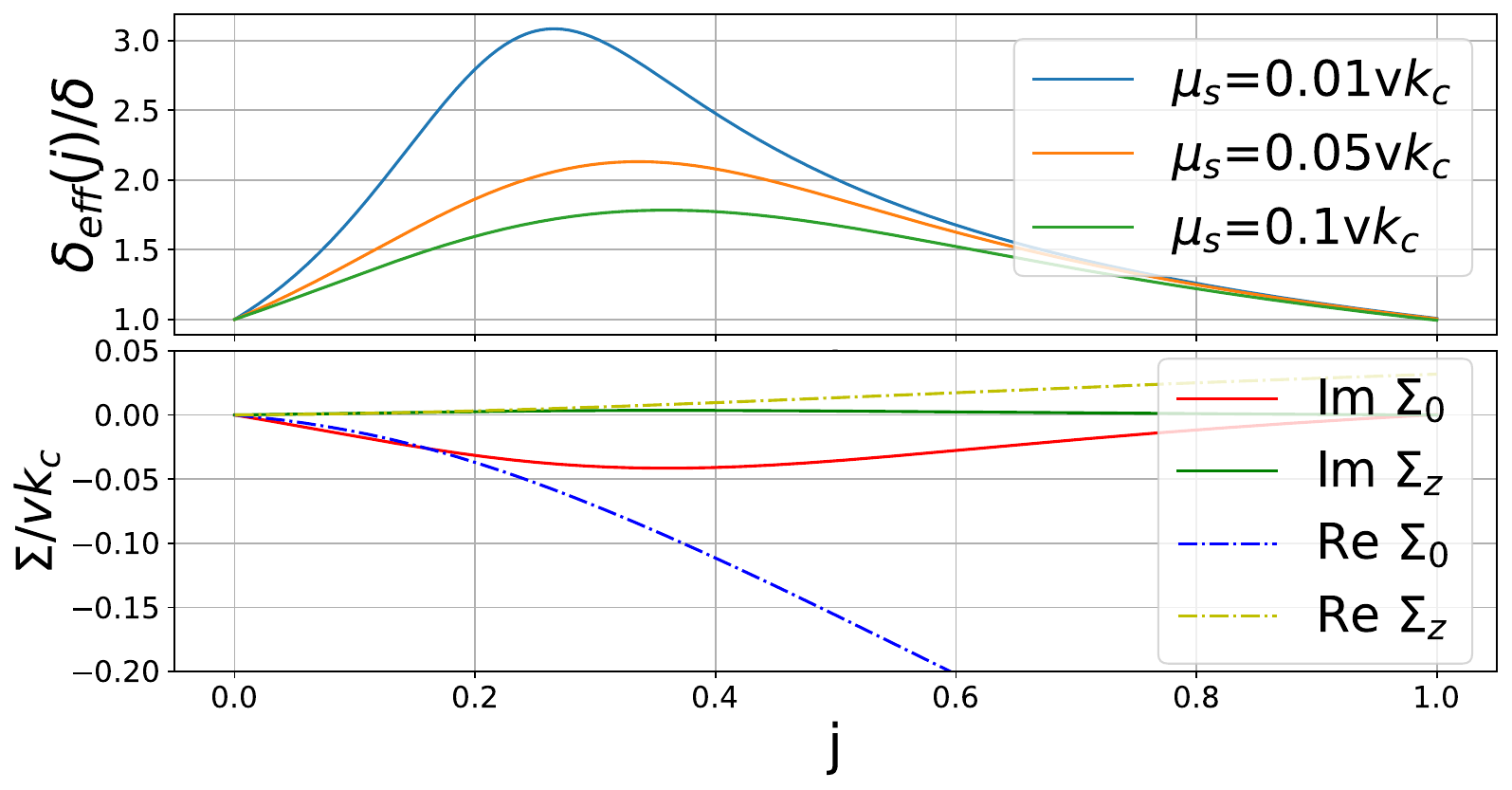}
\caption{Upper figure: the value of the renormalized gap $\delta_{\textrm{eff}}=\delta(\mu_s-\textrm{Re}\,\Sigma_0)/\delta\mu_s$ as a function of the disorder strength $j$ for different values of the surface chemical potential $\mu_s$. Lower figure: the self-energy components as a function of the disorder strength $j$ for $\mu_s=0.1vk_c$. We take $n=5000$, $\mu=0.5m$, $M_z=0.1m$, $A_z=0$ for both figures.}
\label{sj}
\end{figure}

We obtain that in the AFM MTI, the self-energy of the disorder has a non-trivial spin structure $\hat{\Sigma}= \Sigma_0 + \Sigma_z s_z$. We perform a transformation $\Sigma_0 = g_0+\delta g_z, \Sigma_z = \delta g_0+g_z$, and after integration over momentum and summation over $i$ derive from Eqs.~\eqref{born_eq}
\begin{eqnarray}\label{born_series}
g_0^{(n+1)}\!&=&\!\!j(1-\delta^2)\frac{{\tilde\mu}_{\textrm{s}}-g_0^{(n)}}{2}\Xi^{(n)}, \\
\nonumber
g_z^{(n+1)}\!&=&\!\frac12\; j(1-\delta^2)g_z^{(n)}\Xi^{(n)},\\
\Xi^{(n)} = &\ln&\frac{\tilde{v}^2k_c^2}{(\delta^2-1)\left[\left({\tilde\mu}_{\textrm{s}}-g_0^{(n)}\right)^2-g_z^{2(n)}\right]}, \nonumber
\end{eqnarray}
where $k_c$ is a cut-off momentum. The obtained result is equivalent to the self-energy equations for the Dirac Hamiltonian without magnetization~\cite{Akzyanov2023}. Following Ref.~\cite{Akzyanov2023}, we take $g_0^{(0)}=-i0$ and $g_z^{(0)}=0$, which implies that $g_z^{(n)}=0$ and $\Sigma_z^{(n)}=\delta\Sigma_0^{(n)}$. We plot self-energy components in Fig.~\ref{sj}. We see that the disorder generates a real part of the self-energy. The main effect of the disorder is the increase of the surface chemical potential ${\tilde\mu}_{\textrm{s}} \rightarrow {\tilde\mu}_{\textrm{s}}-\textrm{Re}\,\Sigma_0$, which gives rise to the increase of the gap $\delta{\tilde\mu}_{\textrm{s}}$. 

\end{document}